\documentstyle[12pt]{article}

\begin{document}

\title{\begin{flushright}
\begin{small} PUPT-1587 \end{small}
\end{flushright}
Statistical Inference, Occam's Razor and Statistical Mechanics
on The Space of Probability Distributions}

\author{Vijay Balasubramanian\thanks{vijayb@puhep1.princeton.edu} \\  
         {\it Dept. of Physics, Princeton University, Princeton, NJ 08544}}
\date{\today} 

\maketitle

\begin{abstract}
The task of parametric model selection is cast in terms of a
statistical mechanics on the space of probability distributions.
Using the techniques of low-temperature expansions, we arrive at a
systematic series for the Bayesian posterior probability of a model
family that significantly extends known results in the literature.  In
particular, we arrive at a precise understanding of how Occam's Razor,
the principle that simpler models should be preferred until the data
justifies more complex models, is automatically embodied by
probability theory.  These results require a measure on the space of
model parameters and we derive and discuss an interpretation of
Jeffreys' prior distribution as a uniform prior over the distributions
indexed by a family.  Finally, we derive a theoretical index of the
complexity of a parametric family relative to some true distribution
that we call the {\it razor} of the model.  The form of the razor
immediately suggests several interesting questions in the theory of
learning that can be studied using the techniques of statistical
mechanics.
\end{abstract}

\section{Introduction}

In recent years increasingly precise experiments have directed the
interest of biophysicists towards learning in simple neural systems.
The typical context of such learning involves estimation of some
behaviourally relevant information from a statistically varying
environment.  For example, the experiments of de Ruyter and
collaborators have provided detailed measurements of the adaptive
encoding of wide field horizontal motion by the H1 neuron of the
blowfly (\cite{ruyter}).  Under many circumstances the associated
problems of statistical estimation can be fruitfully cast in the
language of statistical mechanics, and the powerful techniques
developed in that discipline can be brought to bear on questions
regarding learning (\cite{potters}).

In this paper we are concerned with a problem that arises frequently
in the context of biophysical and computational learning - the
estimation of parametric models of some true distribution $t$ based on
a collection of data drawn from $t$.  If we are given a particular
family of parametric models (Gaussians, for example) the task of
modelling $t$ is reduced to parameter estimation, which is a
relatively well-understood, though difficult, problem.  Much less is
known about the task of model family selection - for example, how do
we choose between a family of Gaussians and a family of fifty
exponentials as a model for $t$ based on the available data?  In this
paper we will be concerned with the latter problem on which
considerable ink has already been expended in the literature
(\cite{riss84,riss86}, \cite{barron85,clarke,barron},\cite{wallace},
\cite{yamanishi}, \cite{mackay1,mackay2}).

The first contribution of this paper is to cast Bayesian model family
selection more clearly as a statistical mechanics on the space of
probability distributions in the hope of making this important problem
more accessible to physicists.  In this language, a finite dimensional
parametric model family is viewed as a manifold embedded in the space
of probability distributions. The probability of the model family
given the data can be identified with a partition function associated
with a particular energy functional.  The formalism bears a
resemblance to the description of a disordered system in which the
number of data points plays the role of the inverse temperature and in
which the data plays the role of the disordering medium.  Exploiting
the techniques of low temperature expansions in statistical mechanics
it is easy to extend existing results that use Gaussian approximations
to the Bayesian posterior probability of a model family to find
``Occam factors" penalizing complex models
(\cite{clarke,mackay1,mackay2}).  We find a systematic expansion in
powers of $1/N$ where $N$ is the number of data points and identify
terms that encode accuracy, model dimensionality and robustness as
well as higher order measures of simplicity. The subleading terms can
be important when the number of data points is small and represent a
limited attempt to move analysis of Bayesian statistics away from
asymptotics towards the regime of small $N$ that is often biologically
relevant.  The results presented here do not require the true
distribution to be a member of the parametric family under
consideration and the model degeneracies that can threaten analysis in
such cases are dealt with by the method of collective coordinates from
statistical mechanics.  Some connections with the Minimum Description
Length principle and stochastic complexity are discussed
(\cite{barron,clarke,wallace,riss84,riss86}).

In order to perform Bayesian model selection it is necessary to have a
prior distribution on the space of parameters of a model.
Equivalently, we require the correct measure on the phase space
defined by the parameter manifold in the analogue statistical
mechanical problem considered in this paper.  In the absence of
well-founded reasons to pick a particular prior distribution, the
usual prescription is to pick an unbiased prior density that weights
all parameters equally.  However, this prescription is not invariant
under reparametrization and we will argue that the correct prior
should give equal weight to all {\it distributions} indexed by the
parameters.  Requiring all distributions to be a priori equally
likely yields Jeffreys' prior on the parameter manifold, giving a new
interpretation of this choice of prior density (\cite{jeffreys}).

Finally, consideration of the large $N$ limit of the asymptotic
expansion of the Bayesian posterior probability leads us to define the
{\it razor} of a model, a theoretical index of the complexity of a
parametric family relative to a true distribution.  In statistical
mechanical terms, the razor is the quenched approximation of the
disordered system studied in Bayesian statistics.  Analysis of the
razor using the techniques of statistical mechanics can give insights
into the types of phenomena that can be expected in systems that
perform Bayesian statistical inference.  These phenomena include
``phase transitions" in learning  and adaptation to changing
environments.  In view of the length of this paper, applications of
the general framework developed here to specific models relevant to
biophysics will be left to future publications.

\section{Statistical Inference and Statistical Mechanics} 
Suppose we are given a collection of outcomes $E = \{ e_1 \ldots e_N
\}, \, e_i \in X $ drawn independently from a density $t$.  Suppose
also that we are given two parametric families of distributions A and
B and we wish to pick one of them as the model family that we will
use.  The Bayesian approach to this problem consists of computing the
posterior conditional probabilities $\Pr(A | E)$ and $\Pr(B | E)$ and
picking the family with the higher probability.  Let A be parametrized
by a set of real parameters $\Theta = \{\theta_1, \ldots \theta_d \}$.
Then Bayes Rule tells us that:
\begin{equation}                                  
\label{eq:bayes1}                                  
\Pr(A | E) = \frac{\Pr(A)}{\Pr(E)} \int d^d\Theta \, \, 
             w(\Theta) \Pr(E|\Theta) 
\end{equation}                                 
In this expression $\Pr(A)$ is the prior probability of the model
family, $w(\Theta)$ is a prior density on the parameter space and
$Pr(E)$ is a prior density on the $N$ outcome sample space.  The
measure induced by the parametrization of the $d$ dimensional
parameter manifold is denoted $d^d\Theta$.  Since we are interested in
comparing $\Pr(A | E)$ with $\Pr(B | E)$, the prior $\Pr(E)$ is a
common factor that we may omit, and for lack of any better choice we
take the prior probabilities of A and B to be equal and omit them.
Finally, throughout this paper we will assume that the model families
of interest to us have compact parameter spaces.  This condition is
easily relaxed by placing regulators on non-compact parameter spaces,
but we will not concern ourselves with this detail here.

\subsection{Natural Priors or Measures on Phase Space}
\label{sec:priors}
In order to make further progress we must identify the prior density
$w(\Theta)$.  In the absence of a well-motivated prior, a common
prescription is to use the uniform distribution on the parameter space
since this is deemed to reflect complete ignorance (\cite{mackay1}).
In fact, this choice suffers from the serious deficiency that the
uniform priors relative to different parametrizations can assign
different probability masses to the same subset of parameters
(\cite{jeffreys,lee}).  Consequently, if $w(\Theta)$ was uniform in
the parameters, the probability of a model family would depend on the
arbitrary parametrization.  The problem can be cured by making the
much more reasonable requirement that all {\it distributions} rather
than all {\it parameters} are equally likely.\footnote{This applies
the principle of maximum entropy on the invariant space of
distributions rather than the arbitrary space of parameters} In order
to implement this requirement we should give equal weight to all
distinguishable distributions on a model manifold.  However, nearby
parameters index very similar distributions.  So let us ask the
question, ``How do we count the number of distinct distributions in
the neighbourhood of a point on a parameter manifold?''  Essentially,
this is a question about the embedding of the parameter manifold in
the space of distributions.  Points that are distinguishable as
elements of $R^n$ may be mapped to indistinguishable points (in some
suitable sense) of the embedding space.

To answer the question, let $\Theta_p$ and $\Theta_q$ index two
distributions in a parametric family and let $E = \{e_1 \cdots e_N\}$
be drawn independently from one of $\Theta_p$ or $\Theta_q$.  In the
context of model estimation, a suitable measure of distinguishability
can be derived by asking how well we can guess which of $\Theta_p$ or
$\Theta_q$ produced $E$.  Let $\alpha_N$ be the probability that
$\Theta_q$ is mistaken for $\Theta_p$ and let $\beta_N$ be the
probability that $\Theta_p$ is mistaken for $\Theta_q$.  Let
$\beta_N^\epsilon$ be the smallest possible $\beta_N$ given that
$\alpha_N < \epsilon$.  Then Stein's Lemma tells us that
$\lim_{N\rightarrow\infty} (-1/N)\ln{\beta_N^\epsilon} =
D(\Theta_p\|\Theta_q)$ where $D(p\|q) = \int dx \, p(x)
\ln(p(x)/q(x))$ is the relative entropy between the densities $p$ and
$q$ (\cite{cover}).

As shown in Appendix~\ref{sec:count}, the proof of Stein's Lemma shows
that the minimum error $\beta_N^\epsilon$ exceeds a
fixed $\beta^*$ in the region where $\kappa/N \geq
D(\Theta_p\|\Theta_q)$ with $\kappa \equiv -\ln{\beta^*} + \ln(1 -
\epsilon)$.\footnote{This assertion is not strictly true.  See
Appendix~\ref{sec:count} for more details.}  By taking $\beta^*$ close
to $1$ we can identify the region around $\Theta_p$ where the
distributions are not very distinguishable from the one indexed by
$\Theta_p$.  As N grows large for fixed $\kappa$, any $\Theta_q$ in
this region is necessarily close to $\Theta_p$ since
$D(\Theta_p\|\Theta_q)$ attains a minimum of zero when $\Theta_p =
\Theta_q$.  Therefore, setting $\Delta\Theta = \Theta_q - \Theta_p$,
Taylor expansion gives $D(\Theta_p\|\Theta_q) \approx
(1/2)J_{ij}(\Theta_p) \Delta\Theta^i\Delta\Theta^j +
O(\Delta\Theta^3)$ where $J_{ij} = \nabla_{\phi_i} \nabla_{\phi_j}
D(\Theta_p \| \Theta_p + \Phi)|_{\Phi = 0}$ is the Fisher Information.
\footnote{We have assumed that the derivatives with respect to
$\Theta$ commute with expectations taken in the distribution
$\Theta_p$ to identify the Fisher Information with the matrix of
second derivatives of the relative entropy.} (We use the convention
that repeated indices are summed over.)  

In a certain sense, the relative entropy, $D(\Theta_p\|\Theta_q)$,
appearing in this problem is the natural distance between probability
distributions in the context of model selection.  Although it does not
itself define a metric, the Taylor expansion locally yields a
quadratic form with the Fisher Information acting as the metric.  If
we accept $J_{ij}$ as the natural metric, differential geometry
immediately tells us that the reparametrization invariant measure on
the parameter manifold is $d^d\Theta \, \sqrt{\det{J}}$
(\cite{amari85,amari87}).  Normalizing this measure by dividing by
$\int d^d\Theta \sqrt{\det{J}}$ gives the so-called Jeffreys' prior on
the parameters.

A more satisfying explanation of the choice of prior proceeds by
directly counting the number of distinguishable distributions in the
neighbourhood of a point on a parameter manifold.  Define the {\it
volume of indistinguishability} at levels $\epsilon$, $\beta^*$, and
$N$ to be the volume of the region around $\Theta_p$ where $\kappa/N
\geq D(\Theta_p\|\Theta_q)$ so that the probability of error in
distinguishing $\Theta_q$ from $\Theta_p$ is high.
We find to leading order:
\begin{equation}
V_{\epsilon,\beta^*,N} =
\left( \frac{2\pi\kappa}{N} \right)^{d/2} 
\frac{1}{\Gamma(d/2 + 1)}
\frac{1}{\sqrt{\det{J_{ij}(\Theta_p)}}}
\end{equation}
If $\beta^*$ is very close to one, the distributions inside
$V_{\epsilon,\beta^*,N}$ are not very distinguishable and the
Bayesian prior should not treat them as separate distributions.  We
wish to construct a measure on the parameter manifold that reflects
this indistinguishability.  We also assume a principle of
``translation invariance" by supposing that volumes of
indistinguishability at given values of $N$, $\beta^*$ and $\epsilon$
should have the same measure regardless of where in the space of
distributions they are centered. An integration measure reflecting
these principles of indistinguishability and translation invariance
can be defined at each level $\beta^*$, $\epsilon$, and $N$ by
covering the parameter manifold economically with volumes of
indistinguishability and placing a delta function in the center of
each element of the cover. This definition reflects
indistinguishability by ignoring variations on a scale smaller than
the covering volumes and reflects translation invariance by giving
each covering volume equal weight in integrals over the parameter
manifold.  The measure can be normalized by an integral over the
entire parameter manifold to give a prior distribution.  The continuum
limit of this discretized measure is obtained by taking the limits
$\beta^* \rightarrow 1$, $\epsilon\rightarrow 0$ and $N \rightarrow
\infty$.   In this limit the measure counts distributions that are
completely indistinguishable ($\beta^* = 1$)  even in the presence of
an infinite amount of data ($N=\infty$).\footnote{The $\alpha$ and
$\beta$ errors can be treated more symmetrically using the Chernoff
bound instead of Stein's lemma, but we will not do that here.}

To see the effect of the above procedure, imagine a parameter manifold
which can be partitioned into $k$ regions in each of which the Fisher
Information is constant.  Let $J_i$, $U_i$ and $V_i$ be the Fisher
Information, parametric volume and volume of indistintuishability in
the ith region.  Then the prior assigned to the ith volume by the
above procedure will be $P_i = (U_i/V_i) / \sum_{j=1}^k (U_j/V_j) =
U_i \sqrt{\det{J_i}}/\sum_{j=1}^k U_j \sqrt{\det{J_j}}$.  Since all
the $\beta^*$, $\epsilon$ and $N$ dependences cancel we are now free
to take the continuum limit of $P_i$.  This suggests that the prior
density induced by the prescription described in the previous
paragraph is:
\begin{equation}
w(\Theta) = \frac{\sqrt{\det{J(\Theta)}}}{\int d^d\Theta \,
\sqrt{\det{J(\Theta)}}} 
\end{equation}
By paying careful attention to technical difficulties involving sets
of measure zero and certain sphere packing problems, it can be
rigorously shown that the normalized continuum measure on a parameter
manifold that reflects indistinguishability and translation invariance
is $w(\Theta)$ or Jeffreys' prior (\cite{balan}).  In essence, the
heuristic argument above and the derivation in~\cite{balan} show how
to ``divide out'' the volume of indistinguishable distributions on a
parameter manifold and hence give equal weight to equally
distinguishable volumes of distributions.  In this sense, Jeffreys'
prior is seen to be a uniform prior on the {\it distributions} indexed
by a parametric family.

\subsection{Connection With Statistical Mechanics}
\label{sec:conn}
Putting everything together we get the following expression for the
Bayesian posterior probability of a parametric family in the absence
of any prior knowledge about the relative likelihood of the
distributions indexed by the family. 
\begin{equation} 
\Pr(A|E) = \frac{ \int d^d\Theta \sqrt{\det{J}}
         \exp\left[-N \left(\frac{-\ln{\Pr(E|\Theta)}}{N}\right) \right]  } 
           {   \int  d^d\Theta \sqrt{\det{J}}}
\label{eq:bayes2}
\end{equation}
This equation resembles a partition function with a temperature $1/N$
and an energy function $(-1/N)\ln{\Pr(E|\Theta)}$.  The dependence on
the data $E$ is similar to the dependence of a disordered partition
function on the specific set of defects introduced into the system.

The analogy can be made stronger since the strong law of large numbers
says that $(-1/N) \ln\Pr(E|\Theta) = (-1/N) \sum_{i=1}^N
\ln\Pr(e_i|\Theta)$ converges in the almost sure sense to:
\begin{equation}
E_t\left[\frac{-\ln\Pr(e_i|\Theta)}{N}\right]  = \int dx \, t(x)
                 \ln\left(\frac{t(x)}{\Pr(x|\Theta)}\right) - \int dx
                 \, t(x) \ln\left(t(x)\right) = D(t \| \Theta) + h(t)
\label{eq:limitE}
\end{equation}
Here $D(t\|\Theta)$ is the relative entropy between the true
distribution and the distribution indexed by $\Theta$, and $h(t)$ is
the differential entropy of the true distribution that is presumed to
be finite.     With this large $N$ limit in mind we rewrite the
posterior probability in Equation~\ref{eq:bayes2} as the following
partition function:
\begin{equation}
\Pr(A|E) = \frac{\int d^d\Theta \sqrt{J} e^{-N(H_0 + H_d)}}
                {\int d^d\Theta \sqrt{J}}
\label{eq:part}
\end{equation}
where $H_0(\Theta) = D(t\|\Theta)$ and $H_d(E,\Theta) =
(-1/N)\ln\Pr(E|\Theta) - D(t\|\Theta) - h(t)$.
(Equation~\ref{eq:part} differs from Equation~\ref{eq:bayes2} by an
irrelevant factor of $\exp{[-N h(t)]}$). $H_0$ can be regarded as the
``energy'' of the ``state'' $\Theta$ while $H_d$ is the additional
contribution that arises via interaction with the ``defects''
represented by the data.  It is instructive to examine the quenched
approximation to this disordered partition function. (See~\cite{ma} for
a discussion of quenching in statistical mechanical systems.)
Quenching is carried out by taking the expectation value of the energy
of a state in the distribution generating the defects.  In the above
system $E_t[H_d] = 0$ giving the quenched posterior probability:
\begin{equation}
\Pr(A|E)_Q = \frac{\int d^d\Theta \sqrt{J} e^{-N D(t\|\Theta)}}
                  {\int d^d\Theta \sqrt{J}}
\label{eq:quench}
\end{equation}
In Section~\ref{sec:razor} we will see that the logarithm of the
posterior probability converges to the logarithm of the quenched
probability in a certain sense.  This will lead us to regard
the quenched probability as a sort of theoretical index of the
complexity of  a parametric family relative to a given true distribution.

\section{Asymptotic Analysis or Low-Temperature Expansion}
\label{sec:asymp}
Equation~\ref{eq:bayes2} in the previous section represents the full
content of Bayesian model selection.  However, in order to extract
some insight it is necessary to examine special cases.  Let
$\ln{\Pr(E|\Theta)}$ be a smooth function of $\Theta$
that attains a global minimum at $\hat{\Theta}$ and assume that
$J_{ij}(\Theta)$ is a smooth function of $\Theta$ that is positive
definite at $\hat{\Theta}$.  Finally, suppose that
$\hat{\Theta}$ lies in the interior of the compact parameter space and
that the values of local minima are bounded away from the global
minimum by some $b$.\footnote{In Section~\ref{sec:beyond} we will
discuss how to relax these conditions.}  For any given $b$, for
sufficiently large $N$, the Bayesian posterior probabaility will then
be dominated by the neighbourhood of $\hat{\Theta}$ and we can carry
out a low temperature expansion around the saddlepoint at
$\hat{\Theta}$.  

We take the metric on the parameter manifold to be the Fisher
Information since the Jeffreys' prior has the form of a measure
derived from such a metric.  This choice of metric also follows the
work described in~\cite{amari85,amari87}.  We will use $\nabla_\mu$ to
indicate the covariant derivative with respect to $\Theta_\mu$, with a
flat connection for the Fisher Information
metric.\footnote{See~\cite{amari85,amari87} for discussions of
differential geometry in a statistical setting.} Readers who are
unfamiliar with covariant derivatives may read $\nabla_\mu$ as the
partial derivative with respect to $\Theta_\mu$ since we will not be
emphasizing the geometric content of the covariant derivative.

Let $\tilde{I}_{\mu_1\cdots\mu_i} = (-1/N) \nabla_{\mu_1} \cdots
\nabla_{\mu_i} \ln\Pr(E|\Theta) |_{\hat{\Theta}}$ and
$F_{\mu_1\cdots\mu_i} = \nabla_{\mu_1} \cdots \nabla_{\mu_i}
Tr\ln{J_{ij}} |_{\hat{\Theta}}$ where $Tr$ represents the Trace of a
matrix. Writing $(\det{J})^{1/2}$ as $\exp\left[(1/2)Tr\ln{J}\right]$,
we Taylor expand the exponent in the integrand of the Bayesian
posterior around $\hat{\Theta}$, and rescale the integration
variable to $\Phi = N^{1/2}(\Theta - \hat{\Theta})$ to arrive at:
\begin{equation} 
Pr(A|E) = \frac{e^{-\left[       \ln\Pr(E|\hat{\Theta})  - \frac{1}{2}
Tr\ln{J(\hat{\Theta})} \right]}  N^{-d/2}
\int d^d\Phi e^{-((1/2)\tilde{I}_{\mu\nu}\phi^{\mu}\phi^{\nu} +
G(\Phi))}} {\int d^d\Theta \sqrt{\det{J_{ij}}}}
\label{eq:temp1}
\end{equation}
Here $G(\Phi)$ collects the terms that are suppressed by powers of
$N$:
\begin{eqnarray}                                 
\label{eq:gdef}                              
G(\Phi) = & \sum_{i=1}^{\infty} \frac{1}{\sqrt{N^i}} \left[ 
                  \frac{1}{(i+2)!} \tilde{I}_{\mu_1\cdots\mu_{i+2}} 
                                   \phi^{\mu_1}\cdots\phi^{\mu_{(i+2)}} 
                  - \frac{1}{2i!} F_{\mu_1\cdots\mu_i} 
                             \phi^{\mu_1} \cdots \phi^{\mu_i} 
                        \right] \nonumber \\                  
= & \frac{1}{\sqrt{N}} \left[ \frac{1}{3!} \tilde{I}_{\mu_1\mu_2\mu_3}  
\phi^{\mu_1} \phi^{\mu_2} \phi^{\mu_3} - \frac{1}{2} F_{\mu_1}  
\phi^{\mu_1} \right] + \nonumber \\                             
   &        \frac{1}{N}\left[ \frac{1}{4!} \tilde{I}_{\mu_1\cdots\mu_4} 
\phi^{\mu_1}\cdots\phi^{\mu_4} - \frac{1}{2\,2!} F_{\mu_1\mu_2} \phi^{\mu_1} 
\phi^{\mu_2} \right] +    
           O(\frac{1}{N^{3/2}})  
\end{eqnarray}
As before, repeated indices are summed over.  The integral in
Equation~\ref{eq:temp1} may now be evaluated in a series expansion
using a standard trick from statistical mechanics (\cite{id}).  Define a
``source'' $h = \{h_1 \ldots h_d\}$ as an auxiliary variable.  Then it
is easy to verify that:
\begin{equation}
\int d^d\Phi \, e^{-((1/2)\tilde{I}_{\mu\nu}\phi^{\mu}\phi^{\nu} +
G(\Phi))} = e^{- G(\nabla_h)}  
\int d^d\Phi \, e^{-((1/2)\tilde{I}_{\mu\nu}\phi^{\mu}\phi^{\nu} + 
h_\mu \phi^\mu   )} 
\label{eq:gauss}
\end{equation}
where the argument of $G$, $\Phi = (\phi^1\ldots\phi^d)$, has been
replaced by $\nabla_h = \{\partial_{h_1} \ldots \partial_{h_d} \} $
and we assume that the derivatives commute with the integral.  The
remaining obstruction is the compactness of the parameter space.  We
make the final assumption that the bounds of the integration can be
extended to infinity with negligible error since $\hat{\Theta}$ is
sufficiently in the interior or because $N$ is sufficiently large.

Performing the Gaussian integral in Equation~\ref{eq:gauss} and
applying the differential operator $\exp{G(\nabla_h)}$ we find an
asymptotic series in powers of $1/N$.  It turns out to be most useful
to examine $\chi_E(A) \equiv-\ln \Pr(A|E)$.  Defining $V = \int
d^d\Theta \sqrt{\det{J(\Theta)}}$ we find to $O(1/N)$:
\begin{eqnarray}                                 
\label{eq:chie}                                 
\chi_E(A) = & N\left[\frac{-\ln\ Pr(E|\hat{\Theta})}{N}\right]
 + \frac{d}{2} \ln{N}   
- \frac{1}{2}
\ln{\left(\frac{\det{J_{ij}(\hat{\Theta})}}{\det{\tilde{I}_{\mu\nu}(\hat{\Theta})}}\right)}
- 
\ln{\left[\frac{(2\pi)^{d/2}}{V}\right]} + \nonumber \\  
 & \frac{1}{N}  \left\{ \frac{\tilde{I}_{\mu_1\mu_2\mu_3\mu_4}}{4!} \left[
(\tilde{I}^{-1})^{\mu_1\mu_2} (\tilde{I}^{-1})^{\mu_3\mu_4} + \ldots \right] - 
     \frac{F_{\mu_1\mu_2}}{2\,2!} \left[ (\tilde{I}^{-1})^{\mu_1\mu_2} +       
(\tilde{I}^{-1})^{\mu_2\mu_1}  \right]   
\right. - \nonumber \\                   
 &    \frac{\tilde{I}_{\mu_1\mu_2\mu_3}
\tilde{I}_{\nu_1\nu_2\nu_3}}{2!\,3!\,3!} 
 \left[  
(\tilde{I}^{-1})^{\mu_1\mu_2}(\tilde{I}^{-1})^{\mu_3\nu_1}(\tilde{I}^{-1})^{\nu_2\nu_3} 
+ \ldots \right] - \nonumber \\  
 &  \left. \frac{F_{\mu_1} F_{\mu_2}}{2!\,4\,2!\,2!} \left[ 
(\tilde{I}^{-1})^{\mu_1\mu_2} + \ldots    
\right]  + \frac{F_{\mu_1} \tilde{I}_{\mu_2\mu_3\mu_4}}{2!\,2\,2!\,3!}
\left[(\tilde{I}^{-1})^{\mu_1\mu_2} (\tilde{I}^{-1})^{\mu_3\mu_4} + \ldots 
\right]
\right\}
\end{eqnarray}
Terms of higher orders in $1/N$ are easily evaluated with a little
labour, and systematic diagrammatic expansions can be developed
(\cite{id}).  In the next section we will discuss the meaning of
Equation~\ref{eq:chie}

\subsection{Meaning of the Asymptotic Expansion}
\label{sec:meaning}
We can see why the Bayesian posterior measures simplicity and accuracy
of a parametric family by examining Equation~\ref{eq:chie} and noting
that models with larger $\Pr(A|E)$ and hence smaller $\chi_E(A)$ are
better. The $O(N)$ term, $N(-\ln\Pr(E|\hat{\Theta})/N)$, which
dominates asymptotically, is the log likelihood of the data evaluated
at the maximum likelihood point.\footnote{This term is $O(N)$, not
$O(1)$, because $(1/N)\ln\Pr(E|\hat{\Theta})$ approaches a finite
limit at large $N$.}  It measures the accuracy with which the
parametric family can describe the available data.  We will see in
Section~\ref{sec:razor} that for sufficiently large $N$ model families
with the smallest relative entropy distance to the true distribution
are favoured by this term.  The term of $O(N)$ arises from the
saddlepoint value of the integrand in Equation~\ref{eq:bayes2} and
represents the Landau approximation to the partition function.

The term of $O(\ln{N})$ penalizes models with many degrees of freedom and
is a measure of simplicity.  This term arises ``physically" from the
statistical fluctuations around the saddlepoint configuration.
These fluctuations cause the partition function in
Equation~\ref{eq:bayes2} to scale as $N^{-d/2}$ leading to the
logarithmic term in $\chi_E$.  Note that the term of $O(N)$ and
$O(\ln{N})$ have appeared together in the literature as the {\it
stochastic complexity} of a parametric family relative to a collection
of data (\cite{riss84,riss86}).  This definition is justified by
arguing that a family with the lowest stochastic complexity provides
the shortest codes for the data in the limit that $N \rightarrow
\infty$.  Our results suggest that stochastic complexity is merely a
truncation of the logarithm of the posterior probability of a model
family and that adding the subleading terms in $\chi_E$ to the
definition of stochastic complexity would yield shorter codes for
finite $N$.

The $O(1)$ term, which arises from the determinant of quadratic
fluctuations around the saddlepoint, is even more interesting.  The
determinant of $\tilde{I}^{-1}$ is proportional to the volume of the
ellipsoid in parameter space around $\hat{\Theta}$ where the value of
the integrand of the Bayesian posterior is significant.\footnote{If we
fix a fraction $f < 1$ where $f$ is close to 1, the integrand of the
Bayesian posterior will be greater that $f$ times the peak value in an
elliptical region around the maximum.}  The scale for determining
whether $\det \tilde{I}^{-1}$ is large or small is set by the Fisher
Information on the surface whose determinant defines the volume
element.  Consequently the term $\ln(\det J/\det\tilde{I})^{1/2}$ can
be understood as measuring the robustness of the model in the sense
that it measures the relative volume of the parameter space which
provides good models of the data.  More robust models in this sense
will be less sensitive to the precise choice of parameters.  We also
observe from the discussion regarding Jeffreys' prior that the volume
of indistinguishability around $\Theta^*$ is proportional to
$(\det{J})^{-1/2}$.  So the quantity
$(\det{J}/\det{\tilde{I}})^{(1/2)} $ is essentially proportional to
the ratio $V_{large}/V_{indist}$, the ratio of the volume where the
integrand of the Bayesian posterior is large to the volume of
indistinguishability introduced earlier.  Essentially, a model family
is better (more natural or robust) if it contains many distinguishable
distributions that are close to the true.  Related observations have
been made before in~\cite{mackay1,mackay2} and in~\cite{clarke} but
without the interpretation in terms of the robustness of a model
family.

The term $\ln{(2\pi)^d/V}$ can be understood as a preference for
models that have a smaller invariant volume in the space of
distributions and hence are more constrained.  The terms proportional
to $1/N$ are less easy to interpret.  They involve higher derivatives
of the metric on the parameter manifold and of the relative entropy
distances between points on the manifold and the true distribution.
This suggests that these terms essentially penalize high curvatures of
the model manifold, but it is hard to extract such an interpretation
in terms of components of the curvature tensor on the manifold.   It
is worth noting that while terms of $O(1)$ and larger in $\chi_E(A)$
depend at most on the measure (prior distribution) assigned to the
parameter manifold, the terms of $O(1/N)$ depend on the
geometry via the connection coefficients in the covariant derivatives.
For this reason,  the $O(1/N)$ terms are the leading probes of the
effect that the geometry of the space of distributions has on
statistical inference in a Bayesian setting and so it would be very
interesting to analyze them.

Bayesian model family inference embodies Occam's Razor because, for
small $N$, the subleading terms that measure simplicity and robustness
will be important, while for large $N$, the accuracy of the model
family  dominates.

\subsection{Analysis of More General Situations}
\label{sec:beyond}
The asymptotic expansion in Equation~\ref{eq:chie} and the subsequent
analysis were carried out for the special case of a posterior
probability with a single global maximum in the integrand that lay in
the interior of the parameter space.  Nevertheless, the basic insights
are applicable far more generally.  First of all, if the global
maximum lies on the boundary of the parameter space, we can account
for the portion of the peak that is cut off by the boundary and reach
essentially the same conclusions. Secondly, if there are multiple
discrete global maxima, each contributes separately to the asymptotic
expansion and the contributions can be added to reach the same
conclusions.  The most important difficulty arises when the global
maximum is degenerate so that matrix $\tilde{I}$ in
Equation~\ref{eq:chie} has zero eigenvalues.  The eigenvectors
corresponding to these zeroes are tangent to directions in parameter
space in which the value of the maximum is unchanged up to second
order in perturbations around the maximum.  These sorts of
degeneracies are particularly likely to arise when the true
distribution is not a member of the family under consideration, and
can be dealt with by the method of collective coordinates.
Essentially, we would choose new parameters for the model, a subset of
which parametrize the degenerate subspace.  The integral over the
degenerate subspace then factors out of the integral
in Equation~\ref{eq:gauss} and essentially contributes a factor of the
volume of the degenerate subspace times terms arising from the action
of the differential operator $\exp[-G(\nabla_h)]$.  The evaluation of
specific examples of this method in the context of statistical
inference will be left to future publications.

There are situations in which the perturbative expansion in powers of
$1/N$ is invalid.  For example, the partition function in
Equation~\ref{eq:part} regarded as a function of $N$ may have
singularities.  These singularities and the associated breakdown of
the perturbative analysis of this section would be of the utmost
interest since they would be signatures of ``phase transitions" in the
process of statistical inference.  This point will be discussed
further in Section~\ref{sec:biophysics}.

\section{The Razor of  A Model Family}
\label{sec:razor}
The large $N$ limit of the partition function in
Equation~\ref{eq:bayes2} suggests the definition of an ideal
theoretical index of the complexity of a parametric family relative to
a given true distribution.

We know from Equation~\ref{eq:limitE} that
$(-1/N)\ln\left[\Pr(E|\Theta)\right] \rightarrow D(t\|\Theta) + h(t)$
as $N$ grows large.  Now assume that the maximum likelihod estimator
is {\it consistent} in the sense that $\hat{\Theta} =
\arg\max_{\Theta} \ln \Pr(E|\Theta)$ converges in probability to
$\Theta^* = \arg\min_\Theta D(t\|\Theta)$ as $N$ grows
large.\footnote{ In other words, assume that given any neighbourhood
of $\Theta^*$, $\hat{\Theta}$ falls in that neighbourhood with high
probability for sufficiently large $N$.  If the maximum likelihood
estimator is not consistent, statistics has very little to say about
the inference of probability densities.} Also suppose that the log
likelihood of a single outcome $\ln\Pr(e_i|\Theta)$ considered as a
family of functions of $\Theta$ indexed by $e_i$ is {\it
equicontinuous} at $\Theta^*$.\footnote{In other words, given any
$\epsilon > 0$, there is a neighbourhood of $M$ of $\Theta^*$ such
that for every $e_i$ and $\Theta \in M$, $|\ln\Pr(e_i|\Theta) -
\ln\Pr(e_i|\Theta^*)| < \epsilon$.}  Finally, suppose that all
derivatives of $\ln\Pr(e_i|\Theta)$ with respect to $\Theta$ are also
equicontinuous at $\Theta^*$.

Subject to the assumptions in the previous paragraph it is easily
shown that $(-1/N)\ln\Pr(E|\hat{\Theta}) \rightarrow D(t\|\Theta^*) +
h(t)$ as $N$ grows large.  Next, using the covariant derivative with
respect to $\Theta$ defined in Section~\ref{sec:asymp}, let
$\tilde{J}_{\mu_1\cdots\mu_i} = \nabla_{\mu_1} \cdots \nabla_{\mu_i}
D(t\|\Theta) |_{\Theta^*}$.  It also follows that
$\tilde{I}_{\mu_1\cdots\mu_i} \rightarrow
\tilde{J}_{\mu_1\cdots\mu_i}$ (\cite{balan}). Since the terms in the
asymptotic expansion of $(1/N)(\chi_E - N h(t))$
(Equation~\ref{eq:chie}) are continuous functions of
$\ln\Pr(E|\Theta)$ and its derivatives, they individually converge to
limits obtained by replacing each $\tilde{I}$ by $\tilde{J}$ and
$(-1/N)\ln\Pr(E|\hat{\Theta})$ by $D(t\|\Theta^*) + h(t)$.  Define
$(-1/N)\ln R_N(A)$ to be the sum of the series of limits of the
individual terms in the asymptotic expansion of $(1/N)(\chi_E - N
h(t))$:
\begin{equation}
\frac{-\ln R_N(A)}{N} = D(t \|\Theta^*)
 + \frac{d}{2\,N}\ln{N}
-\frac{1}{2\,
N}\ln\left[\frac{\det{J_{ij}(\hat{\Theta})}}{\det{\tilde{J}_{\mu\nu}(\hat{\Theta})}}\right]
- \frac{1}{N}\ln\left[\frac{(2\pi)^{d/2}}{V}\right] 
+ O(\frac{1}{N^2})
\label{eq:chin}
\end{equation}
This formal series of limits can be resummed to obtain:
\begin{equation}
R_N(A) = \frac{\int d^d\Theta \sqrt{J} e^{-N D(t\|\Theta)}}
              {\int d^d\Theta \sqrt{J}}
\label{eq:razor}
\end{equation}
We have encountered $R_N(A)$ before in Section~\ref{sec:conn} as the
quenched approximation to the partition function in
Equation~\ref{eq:bayes2}.  $R_N(A)$ will be called the {\it razor} of the
model family $A$.

The razor, $R_N(A)$, is a theoretical index of the complexity of the
model family $A$ relative to the true distribution $t$ given $N$ data
points.  In a certain sense, the razor is the ideal quantity that
Bayesian methods seek to estimate from the data avaliable in a given
realization of the model inference problem.  Indeed, the quenched
approximation to the Bayesian partition function consists precisely of
averaging over the data in different realizations.  The terms in the
expansion of the log razor in Equation~\ref{eq:chin} are the ideal
analogues of the terms in $\chi_E$ since they arise from derivatives
of the relative entropy distance between distributions indexed by the
model family and the true distribution.  The leading term 
tells us that for sufficiently large $N$, Bayesian inference picks the
model family that comes closest to the true distribution in relative
entropy.  The subleading terms have the same interpretations as the
terms in $\chi_E$ discussed in the previous section, except that they
are the ideal quantities to which the corresponding terms in $\chi_E$
tend when enough data is available.

The razor is useful when we know the true distribution as well as the
model families being used by a particular system and we wish to
analyze the expected behaviour of Bayesian inference.  It is also
potentially useful as a tool for modelling and analysis of the general
types of phenomena that can occur in Bayesian inference - different
relative entropy distances $D(t\|\Theta)$ can yield radically
different learning behaviours as discussed in the next section.  The
razor is considerably easier to analyze than the full Bayesian
posterior probability since the quenched approximation to
Equation~\ref{eq:bayes2} given in Equation~\ref{eq:razor} defines a
statistical mechanics on the space of distributions in which the
``disorder'' has been averaged out.  The tools of statistical
mechanics can then be straightforwardly applied to a system with 
temperature $1/N$ and energy function $D(t\|\Theta)$.

\section{Biophysical Relevance and Some Open Questions}
\label{sec:biophysics}
The general framework described in this paper is relevant to
biophysics if we believe that neural systems optimize their
accumulation of information from a statistically varying environment.
This is likely to be true in at least some circumstances since an
organism derives clear advantages from rapid and efficient detection
and encoding of information.  For example, see the discussions of
Bialek and Atick of neural signal processing systems that approach
physical and information theoretic limits (\cite{bialek,atick}).  A
creature such as a fly is faced with the problem of estimating the
statistical profile of its environment from the small amount of data
available at its retina.  The general formalism presented in this
paper applies to such problems and an optimally designed fly would
implement the formalism subject to the constraints of its biological
hardware.   In this section we will discuss several interesting
questions in the theory of learning that can be discussed effectively
in the statistical mechanical language introduced here.

First of all, consider the possibility of ``phase transitions'' in the
disordered partition function that describes the Bayesian posterior
probability or in the quenched approximation defining the razor.
Phase transitions arise from a competition between entropy and energy
which, in the present context, is a competition between simplicity and
accuracy.  We should expect the existence of systems in which
inference at small $N$ is dominated by ``simpler'' and more ``robust''
saddlepoints whereas at large $N$ more ``accurate'' saddlepoints are
favoured.  As discussed in Section~\ref{sec:meaning}, the
distributions in the neighbourhood of ``simpler'' and more ``robust''
saddlepoints are more concentrated near the true.\footnote{In
Section~\ref{sec:meaning} we have discussed how Bayesian inference
embodies Occam's razor by penalizing complex families until the data
justifies their choice.  Here we are discussing Occam's razor for
choice of saddlepoints {\it within} a given family.}  The transitions
between regimes dominated by these different saddlepoints would
manifest themselves as singularities in the perturbative methods 
that led to the asymptotic expansions for $\chi_E$ and $\ln{R_N(A)}$.

The phase transitions discussed in the previous paragraph are
interesting even when the task at hand is not the comparison of model
families, but merely the selection of parameters for a given family.
In Section~\ref{sec:meaning} we have interpreted  the terms of $O(1)$ 
in $\chi_E$ as measurements of the ``robustness'' or ``naturalness''
of a model.   These robustness terms can be evaluated at different
saddlepoints of a given model and a more robust point may be
preferable at small $N$ since the parameter estimation would then be
less sensitive to fluctuations in the data.

So far we have concentrated on the behaviour of the Bayesian posterior
and the razor as function of the number of data points.  Instead, we
could ask how they behave when the true distribution is changed.  For
example, this can happen in a biophysical context if the environment
sensed by a fly changes when it suddenly finds itself indoors.  In
statistical mechanical terms, we wish to know what happens when the
energy of a system is time-dependent.  If the change is abrupt, the
system will dynamically move between equilibria defined by the energy
functions before and after the change.  If the change is very slow we
would expect adaptation that proceeds gradually.  In the language of
statistical inference, these adaptive processes correspond to learning
of changes in the true distribution.

A final question that has been touched on, but not analyzed, in this
paper is the influence of the geometry of parameter manfiolds on
statistical inference.  As discussed in Section~\ref{sec:meaning},
terms of $O(1/N)$ and smaller in the asymptotic expansions of the
log Bayesian posterior and the log razor depend on details of 
the geometry of the parameter manifold.  It would be very interesting
to understand the precise meaning of this dependence.

\section{Conclusion}
In this paper we have cast parametric model selection as a disordered
statistical mechanics on the space of probability distributions.  A
low temperature expansion was used to develop the asymptotics of
Bayesian methods beyond the analyses avaliable in the literature and
it was shown that Bayesian methods for model family inference embody
Occam's razor.  While reaching these results, we derived and discussed
a novel interpretation of Jeffreys' prior density as the uniform prior
on the probability distributions indexed by a parametric family.  By
considering the large $N$ limit and the quenched approximation of the
disordered system implemented by Bayesian inference, we derived the
{\it razor}, a theoretical index of the complexity of a parametric
family relative to a true distribution.  Finally, in view of the
analogue statistical mechanical interpretation, we discussed various
interesting phenomena that should be present in systems that perform
Bayesian learning.  It is easy to create models that display these
phenomena simply by considering families of distributions for which
$D(t\|\Theta)$ has the right structure.  It would be interesting to
examine models of known biophysical relevance to see if they exhibit
such effects, so that experiments could be carried out to verify their
presence or absence in the real world.  In view of the length of the
present paper, this project will be left to a future publication.

\section{Acknowledgements}
I would like to thank Kenji Yamanishi for several fruitful
conversations.  I have also had useful discussions and correspondence
with Steve Omohundro, Erik Ordentlich, Don Kimber, Phil Chou and Erhan
Cinlar.  Finally, I am grateful to Curt Callan for supporting and
encouraging this project and to Phil Anderson for helping with travel
funds to the 1995 Workshop on Maximum Entropy and Bayesian Methods.
This work was supported in part by DOE grant DE-FG02-91ER40671.

\appendix
\section{Measuring Indistinguishability of Distributions}
\label{sec:count}

Let us take $\Theta_p$ and $\Theta_q$ to be points on a parameter
manifold.  Since we are working in the context of density estimation a
suitable measure of the distinguishability of $\Theta_p$ and
$\Theta_q$ should be derived by taking $N$ data points drawn from
either $p$ or $q$ and asking how well we can guess which distribution
produced the data.  If $p$ and $q$ do not give very distinguishable
distributions, they should not be counted separately since that would
count the same distribution twice.

   Precisely this question of distinguishability is addressed in the
classical theory of hypothesis testing.  Suppose $\{e_1 \ldots e_N \}
\in E^N$ are drawn iid from one of $f_1$ and $f_2$ with $D(f_1\|f_2) <
\infty$. Let $A_N \subseteq E^N$ be the acceptance region for the
hypothesis that the distribution is $f_1$ and define the error
probabilities $\alpha_N = f_1^N(A_N^C)$ and $\beta_N = f_2^N(A_N)$.
($A_N^C$ is the complement of $A_N$ in $E^N$ and $f^N$ denotes the
product distribution on $E^N$ describing $N$ iid outcomes drawn from
$f$.)  In these definitions $\alpha_N$ is the probability that $f_1$
was mistaken for $f_2$ and $\beta_N$ is the probability of the
opposite error.  Stein's Lemma tells us how low we can make $\beta_N$
given a particular value of $\alpha_N$.  Indeed, let us define
$\beta_N^\epsilon = \min_{A_N \subseteq E^N \, , \, \alpha_N \leq
\epsilon}  \beta_N$.
Then Stein's Lemma tells us that:                                 
\begin{equation} 
\lim_{\epsilon \rightarrow 0} \lim_{N \rightarrow \infty} \frac{1}{N}
\ln \beta_N^\epsilon = - D(f_1 \| f_2)
\end{equation}
By examining the proof of Stein's Lemma (\cite{cover}) we find that
for fixed $\epsilon$ and sufficiently large $N$ the optimal choice of
decision region places the following bound on $\beta_N^\epsilon$:
\begin{equation} 
-D(f_1\|f_2) - \delta_N + \frac{\ln{(1 - \alpha_N)}}{N} \leq  
\frac{1}{N} \ln\beta_N^\epsilon \leq 
-D(f_1\|f_2) + \delta_N + \frac{\ln{(1 - \alpha_N)}}{N}
\end{equation}
where $\alpha_N < \epsilon$ for sufficiently large $N$.  The
$\delta_N$ are any sequence of positive constants that satisfy the
property that:
\begin{equation}  
\label{eq:deltas} 
\alpha_N = f_1^N(|\frac{1}{N}\sum_{i=1}^N
\ln{\frac{f_1(e_i)}{f_2(e_i)}} - D(f_1\|f_2) | > \delta_N) \leq
\epsilon
\end{equation}         
for all sufficiently large $N$. 
Now $(1/N) \sum_{i=1}^N \ln(f_1(e_i)/f_2(e_i))$ converges to
$D(f_1\|f_2)$ by the law of large numbers since $D(f_1\|f_2) =
E_{f_1}(\ln(f_1(e_i)/f_2(e_i))$.  So, for any fixed $\delta$ we
have:
\begin{equation}         
\label{eq:deltas2}    
f_1^N(|\frac{1}{N}\sum_{i=1}^N \ln{\frac{f_1(e_i)}{f_2(e_i)}} -
D(f_1\|f_2) | > \delta) < \epsilon
\end{equation}      
for all sufficiently large N.  For a fixed $\epsilon$ and a fixed $N$
let $\Delta_{\epsilon,N}$ be the collection of $\delta > 0$ which
satisfy Equation~\ref{eq:deltas2}.  Let $\delta_{\epsilon N}$ be the
infimum of the set $\Delta_{\epsilon,N}$.  Equation~\ref{eq:deltas2}
guarantees that for any $\delta >0$, for any sufficiently large $N$,
$0 < \delta_{\epsilon N} < \delta$.  We conclude that
$\delta_{\epsilon N}$ chosen in this way is a sequence that converges
to zero as $N \rightarrow \infty$ while satisfying the condition in
Equation~\ref{eq:deltas} which is necessary for proving Stein's Lemma.

We will now apply these facts to the problem of distinguishability of
points on a parameter manifold.   Let $\Theta_p$ and $\Theta_q$ index
two distributions on a parameter manifold and  suppose that we are
given $N$ outcomes generated independently from one of them.  We are
interested in using Stein's Lemma to determine how distinguishable
$\Theta_p$  and $\Theta_q$ are.   By Stein's Lemma: 

\begin{equation}       
\label{eq:st1}       
-D(\Theta_p\|\Theta_q) - \delta_{\epsilon N}(\Theta_q) + \frac{\ln(1 -
\alpha_N)}{N}         \leq       
\frac{\beta_N^\epsilon(\Theta_q)}{N}        \leq       
-D(\Theta_p\|\Theta_q) + \delta_{\epsilon N}(\Theta_q) + \frac{\ln(1 -
\alpha_N)}{N}        
\end{equation}       
where we have written $\delta_{\epsilon N}(\Theta_q)$ and
$\beta_N^\epsilon(\Theta_q)$ to emphasize that these quantities are
functions of $\Theta_q$ for a fixed $\Theta_p$.  Let $A =
-D(\Theta_p\|\Theta_q) + (1/N)\ln(1- \alpha_N)$ be the average of the
upper and lower bounds in Equation~\ref{eq:st1}.  Then $A \geq
-D(\Theta_p\|\Theta_q) + (1/N) \ln(1 - \epsilon)$ because the
$\delta_{\epsilon N}(\Theta_q)$ have been chosen to satisfy
Equation~\ref{eq:deltas}.  We now define the set of distributions $U_N
= \{\Theta_q : -D(\Theta_p\|\Theta_q) + (1/N) \ln(1 - \epsilon) \geq
(1/N) \ln{\beta^*} \}$ where $1 > \beta^* > 0$ is some fixed
constant. Note that as $N \rightarrow \infty$, $D(\Theta_p\|\Theta_q)
\rightarrow 0$ for $\Theta_q \in U_N$.  We want to show that $U_N$ is
a set of distributions which cannot be very well distinguished from
$\Theta_p$.  The first way to see this is to observe that the average
of the upper and lower bounds on $\ln{\beta^\epsilon_N}$ is greater
than or equal to $\ln\beta^*$ for $\Theta_q \in U_N$.  So, in this
loose, average sense, the error probability $\beta_N^\epsilon$ exceeds
$\beta^*$ for $\Theta_q \in U_N$.  

More carefully, note that $(1/N) \ln(1 - \alpha_N) \geq (1/N) \ln(1 -
\epsilon)$ by choice of the $\delta_{\epsilon N}(\Theta_q)$.  So,
using Equation~\ref{eq:st1} we see that $(1/N)
\ln{\beta_N^\epsilon(\Theta_q)} \geq (1/N) \ln{\beta^*} -
\delta_{\epsilon N}(\Theta_q)$.  Exponentiating this inequality we
find that:
\begin{equation}     
\label{eq:expbound}  
1       
\geq        
\left[ \beta_N^\epsilon(\Theta_q) \right]^{(1/N)}        
\geq       
(\beta^*)^{(1/N)} \, e^{-\delta_{\epsilon N}(\Theta_q)}       
\end{equation}       
The significance of this expression is best understood by considering
parametric families in which, for every $\Theta_q$, $X_q(e_i) =
\ln(\Theta_p(e_i)/\Theta_q(e_i))$ is a random variable with finite
mean and bounded variance, in the distribution indexed by
$\Theta_p$. In that case, taking $b$ to be the bound on the variances,
Chebyshev's inequality says that: 
\begin{equation}     
\Theta_p^N\left(|\frac{1}{N} \sum_{i=1}^N X_q(e_i) \, - \,
D(\Theta_p\|\Theta_q) | > \delta   
\right)        
\leq        
\frac{Var(X)}{\delta^2 \, N}        
\leq        
\frac{b}{\delta^2 \, N}       
\end{equation}       
In order to satisy $\alpha_N \leq \epsilon$ it suffices to choose
$\delta = (b/N\epsilon)^{1/2}$.  So, if the bounded variance condition
is satisfied, $\delta_{\epsilon N}(\Theta_q) \leq (b/N\epsilon)^{1/2}$
for any $\Theta_q$ and therefore we have the limit $\lim_{N
\rightarrow \infty} \sup_{\Theta_q \in U_N} \delta_{\epsilon
N}(\Theta_q) = 0$.  Applying this limit to Equation~\ref{eq:expbound}
we find that:
\begin{equation}       
\label{eq:asstein}       
1        
\geq       
\lim_{N \rightarrow \infty} \inf_{\Theta_q \in U_N}
\left[\beta_N^\epsilon(\Theta_q)         
\right]^{(1/N)} \geq 1 \times \lim_{N \rightarrow \infty}
\inf_{\Theta_q \in U_N}        
e ^{-\delta_{\epsilon N}(\Theta_q)} = 1        
\end{equation}       
In summary we find that $\lim_{N \rightarrow \infty} \inf_{\Theta_q
\in U_N} [ \beta_N^\epsilon(\Theta_q) ]^{(1/N)} = 1$.  This is to be
contrastd with the behaviour of $\beta_N^\epsilon(\Theta_q)$ for any
{\it fixed} $\Theta_q \neq \Theta_p$ for which $\lim_{N \rightarrow
\infty} [\beta_N^\epsilon(\Theta_q)]^{(1/N)} = \exp{-
D(\Theta_p\|\Theta_q)} < 1 $.  We have essentially shown that the sets
$U_N$ contain distributions that are not very distinguishable from
$\Theta_p$.  The smallest one-sided error probability
$\beta_N^\epsilon$ for distinguishing between $\Theta_p$ and $\Theta_q
\in U_N$ remains essentially constant leading to the asymptotics in
Equation~\ref{eq:asstein}. 
       
Define $\kappa \equiv -\ln\beta^* + \ln(1-\epsilon)$ so that we can
summarize the region $U_N$ of high probability of error $\beta^*$ at
fixed $\epsilon$ as $\kappa/N \geq D(\theta_p\|\theta_q)$.
In this region, the distributions are indistinguishable from $\Theta_p$
with error probabilities $\alpha_N \leq \epsilon$ and
$(\beta_N^\epsilon)^{(1/N)} \geq (\beta^*)^{(1/N)}
\exp{-\delta_{\epsilon N}}$.

\end{document}